# Proton-fluence dependent magnetic properties of exfoliable quasi-2D van der Waals Cr$_2$Si$_2$Te$_6$ magnet

Hector Iturriaga[1#], Ju Chen[2#], Jing Yang[2], Luis M. Martinez[1,3], Martin Kirk[2,4,5&], Lin Shao[6], Y. Liu[7,$], C. Petrovic[7], Srinivasa R. Singamaneni[1*]

[1]*Department of Physics, The University of Texas at El Paso, El Paso, TX 79968, USA*

[2]*Department of Chemistry and Chemical Biology, The University of New Mexico, MSC03 2060, 1 University of New Mexico, Albuquerque, New Mexico 87131-0001, United States,*

[3]*Center for Integrated Nanotechnologies, Los Alamos National Laboratory, Los Alamos, NM 87545*

[4]*The Center for High Technology Materials, The University of New Mexico, Albuquerque, New Mexico 87106, United States*

[5]*Center for Quantum Information and Control (CQuIC), The University of New Mexico, Albuquerque, New Mexico 87131-0001, United States.*

[6]*Department of Nuclear Engineering, Texas A&M University, College Station, TX 77845, USA*

[7]*Condensed Matter Physics and Materials Science Department, Brookhaven National Laboratory, Upton, New York 11973, USA*

[$]*Present address: Los Alamos National Laboratory, Los Alamos, New Mexico 87545, USA*

## Abstract

The discovery of long-range magnetic ordering in atomically thin materials catapulted the van der Waals (vdW) family of compounds into an unprecedented popularity. In particular, with a current push in space exploration, it is beneficial to study how the properties of such materials evolve under proton irradiation. Owing to their robust intra-layer stability and sensitivity to external perturbations, these materials provide excellent opportunities for studying proton irradiation as a non-destructive tool for controlling their magnetic properties. Specifically, the exfoliable Cr$_2$Si$_2$Te$_6$ (CST) is a ferromagnetic semiconductor with the Curie temperature (T$_C$) of ~32 K. Here, we have investigated the magnetic properties of CST upon proton irradiation as a function of fluence (1 x 10$^{15}$, 5 x 10$^{15}$, 1 x 10$^{16}$, 5 x 10$^{16}$, and 1 x 10$^{18}$ H$^+$/cm$^2$) by employing variable-temperature, variable-field magnetization measurements coupled with electron paramagnetic resonance (EPR) spectroscopy and detail how the magnetization, magnetic anisotropy and EPR spectral parameters vary as a function of proton fluence across the magnetic phase transition. While the T$_C$ remains constant as a function of proton fluence, we observed that the saturation magnetization and magnetic anisotropy diverge at the proton fluence of 5 x 10$^{16}$ H$^+$/cm$^2$, which is prominent in the ferromagnetic phase, in particular. This work demonstrates that proton irradiation is a feasible method for modifying the magnetic properties and local magnetic interactions of vdWs crystals, which represents a significant step forward in the design of future spintronic and magneto-electronic applications.

*srao@utep.edu

&mkirk@unm.edu

#equally contributed

**Introduction and Motivation**

  Pioneering transition metal-based vdW magnets such as $CrI_3$ and $Cr_2Ge_2Te_6$ have been successfully studied at the single-layer limit, displaying novel and exciting long-range magnetic ordering at the 2-dimensional level. This offers hope for the creation of ultrathin magnetic data storage devices, spintronic applications, and other cost-effective magnetic devices. [1–5] Of particular importance, with renewed interest in space exploration, it is beneficial to investigate the behavior of these materials under the effects of proton irradiation. In space, where proton irradiation is abundant, bombardment by such particles can quickly modify their electronic and magentic properties. Thus, the robust intralayer properties of these compounds make them excellent candidates for studying the limits of proton irradiation on such functional materials. Particularly, though not exclusively, in the case of ternary transition metal-based vdW chalcogenides such as $Fe_{2.7}GeTe_2$ (FGT) and $Mn_3Si_2Te_6$ (MST), our group has found that proton irradiation serves as a non-destructive tool for controlling the magnetic properties of these compounds. [6–11] These compounds owe their enhanced magnetization after irradiation to their strong intralayer bonds, robust magnetic character, and sensitivity to external perturbations.

  Here, we explore the effects of non-destructive proton irradiation on the related compound $Cr_2Si_2Te_6$ (CST). Paired with its layer-dependent magnetism, proton-irradiated CST could lead to breakthroughs in the creation of functional 2D magnets, novel heterostructures, and new high-speed magnetic devices. In this ternary chromium chalcogenide, the ferromagnetic ordering temperature is only 32 K, making it unsuitable for consumer applications. However, its magnetic properties are stable even under extreme conditions; hence, proton irradiation may be a suitable approach to enhance the properties of CST without destroying it. For this reason, we have investigated the evolution of the Curie temperature ($T_C$), saturation magnetization, and magnetocrystalline anisotropy in CST crystals irradiated with proton doses of up $1 \times 10^{18}$ $(H^+)/cm^2$ by employing variable-temperature, variable-field magnetization measurements coupled with electron paramagnetic resonance (EPR) spectroscopy.

**Experimental Details**

The synthesis of mm-sized CST crystals was conducted as previously reported with the self-flux method starting with Cr powder and Si and Te pieces. [12] The crystals were then irradiated with a 2 MeV proton beam in 1.7 MV Tandetron Accelerator following closely the procedure discussed. [6-9] The energy of the beam is kept low enough to avoid damage to the crystal. In addition, the weak beam current of 100 nA, beam spot size of 6mm x 6mm rastered over an area of 1.2 cm x 1.2 cm, and magnetic bending filters ensure lateral beam uniformity, minimal heating, and eliminate carbon contamination. The projected implantation depth is approximately 35 microns according to the damage profile, which shows a uniform distribution of H from the surface down to approximately 30 microns from the surface, at which point there is a noticeable increase. The fluences selected for this study are 0 (referred to as pristine in this work), $1 \times 10^{15}$, $5 \times 10^{15}$, $1 \times 10^{16}$, $5 \times 10^{16}$, and $1 \times 10^{18}$ H$^+$/cm$^2$.

The magnetic properties of the crystals were characterized using a Quantum Design Magnetic Property Measurement System 3 (MPMS 3) Superconducting QUantum Interference Device/Vibrating Sample Magnetometer (SQUID/VSM) hybrid system. DC magnetic moment data were extracted using the VSM option throughout all temperature and field ranges. The measurements were conducted in a temperature range of 2 K – 300 K and at magnetic fields up to 30 kOe in both polarities. Samples were loaded separately into the MPMS by wrapping them in Teflon tape to avoid contamination from sample holders or from contact with the instrument chamber. In the case of magnetic moment anisotropy measurements, we used both quartz and brass sample holders, which allow for in-plane (where the crystallographic ab-plane is parallel to the applied field) and out-of-plane (where the c-axis is parallel to the applied field) magnetization measurements. Contributions to the magnetic moment from the sample holders was corrected using the MPMS 3 MultiVu software.

Temperature-dependent X-band (~9.49 GHz) EPR spectroscopic studies were also performed for each sample. Samples were individually wrapped with Teflon tape and mounted onto a sample holder assembly such that the c-axis of the crystal was oriented parallel to the applied magnetic field. Measurements were performed by first setting the temperature and allowing the samples to reach thermal equilibrium, and then the scanning the magnetic field between 0 G to 7000 G. To avoid the effects of remanent magnetization between temperature points, the magnetic field was removed, the sample was warmed above its Curie temperature, and finally cooled down to the next temperature point. Measurements were performed at select temperatures between 10 K and 77 K.

**Results and Discussion**

Figure 1a shows a plot of magnetization vs. temperature at various proton fluences. The data were collected along the out-of-plane (H//c) direction with a cooling field of 1 kOe. These field cooled (FC) measurements were conducted by cooling the sample from room temperature down to 10 K under an applied magnetic field. The sample was then allowed to reach thermal equilibrium before the magnetic field was set to 500 Oe, and then the temperature was increased incrementally back to 300 K. For a better visualization of the ferromagnetic phase of CST, only the range from 10 K – 100 K is shown in Figure 1, since the samples are paramagnetic beyond 50 K. As reflected in this figure, a sharp ferromagnetic to paramagnetic transition is noticed for all the samples regardless of proton fluence. Estimation of the Curie temperature ($T_C$) via the first derivative of the magnetization with respect to the temperature (Figure 1b) yields a value of approximately 36 K, which agrees well with literature reports. [12–15] In addition, the low temperature behavior shows that there is a clear change in magnetization after irradiation. Unexpectedly, the magnetization of the 5 x $10^{15}$ H$^+$/cm$^2$ fluence (orange, upright triangles) slightly exceeds the magnetization of the pristine sample at 10 K, before rapidly dropping below the magnetization of the pristine sample at 12 K. For all other proton fluences, we observed a clear decrease in the magnetization. We believe that there is a small Curie-Weiss component induced by the proton irradiation as all the curves below $T_C$ turn upwards at lowest temperatures.

Particularly, at 10 K, the sample irradiated with 5 x $10^{16}$ H$^+$/cm$^2$ (Figure1a, green rhombus) displays a magnetization of 4.5 emu/g, a remarkable 40% decrease from the pristine magnetization of 7.5 emu/g. The sample with 5 x $10^{16}$ H$^+$/cm$^2$ fluence shows the most dramatic decrease in magnetization, hence the comparison in the magnetization between this compound and the pristine compound will be emphasized for the following discussion of anisotropic temperature-dependent measurements. These measurements were also performed along the in-plane direction (H//ab) to gather information about the magnetocrystalline nature of the samples. Figure 2 details a comparison of the H//c (solid symbols) and H//ab (open symbols) magnetization vs. temperature curves for both FC (cyan symbols) and ZFC (navy blue symbols) measurements. ZFC measurements were performed by cooling the sample in the absence of a magnetic field. As expected, the magnetization is higher in the H//c direction, as this is the magnetic easy axis of CST and provides the evidence for a strong perpendicular magnetocrystalline anisotropy in the system. [12–16] This is also observed for the irradiated samples, providing evidence that the easy axis remains stable despite a decrease in the magnetization. The H//c measurement for the irradiated sample (Figure 2b) shows a large bifurcation between the FC and ZFC curves at low temperatures that is not present in the pristine magnetization (Figure 2a), suggesting alterations to the magnetic anisotropy of irradiated CST. In contrast, the H//ab magnetization data is similar in both samples, indicating that the applied field is too small to accurately detect any anisotropic effects in this direction. It may also be due to the spin glass behavior.

Figure 3 shows the isothermal magnetic field-dependent magnetization at all proton fluences as a function of temperatures both below (Figures 3a-d) and above (figures 3e and f) $T_C$ (~ 36 K). Samples were cooled from room temperature to the target measurement temperature and then allowed to reach equilibrium in the absence of a magnetic field. For these measurements, the samples were subjected to a 5-quadrant field sweep, where the field is ramped up from 0 Oe, up to the maximum field of +30 kOe, and then incremented back through 0 Oe and down to -30 kOe, before finally increasing the field back to its maximum value. For isotherms below $T_C$, the system is warmed above $T_C$ to its paramagnetic state in the absence of a magnetic field upon completing the measurement and cooled to the next target temperature to avoid remanence effects. The measurements were performed in both the H//c (Figures 3a, c, and d, solid symbols) and H//ab (Figures 3b, d, and f, open symbols) directions, which allows us to qualitatively and quantitatively analyze the magnetocrystalline anisotropy of irradiated CST. From these data, we also extract information regarding the effect proton irradiation on the saturation magnetization and coercivity of CST.

The hysteresis curves presented in Figures 3a – d offer clear proof of the ferromagnetic phase of CST at temperatures below $T_C$. Even after irradiation, all of the samples exhibit typical ferromagnetic behavior. We also observe negligible coercivity at all proton fluences. In comparison to the H//ab data (Figure 3b), the H//c saturation field is markedly lower, and this provided additional evidence in support of CST maintaining its perpendicular magnetic anisotropy with an H//c easy axis. Conversely, the H//ab magnetization saturates only at higher fields and shows a more gradual onset of saturation. The most obvious change that occurs after irradiation is a reduction in the saturation magnetization.

Regarding the 5 x $10^{16}$ H+/cm$^2$ fluence data, the magnetization decreases by over 30%, and this supports our initial observation in the temperature-dependent magnetization measurements. Figure 4 shows the saturation magnetization M$_{sat}$ for all fluences as a function of temperature. The values of M$_{sat}$ are taken from the H//c hysteresis curves at the maximum field. This plot reveals that the reduction in magnetization is not proportional to the number of protons per unit area. In fact, the median fluence of 1 x $10^{16}$ H+/cm$^2$ (yellow, down triangle) shows lower values of magnetization than the highest fluence of 1 x $10^{18}$ H+/cm$^2$ (blue, left triangle). Moreover, at a fluence of 5 x $10^{16}$ H+/cm$^2$, the data shows the lowest magnetization despite the sample not being subjected to the highest proton fluence. We observe a non-monotonic relationship between the proton fluence and the magnetization (Figure 4a). As a result, we cannot simply attribute the decrease in magnetization to a destruction of the crystal since the 1 x $10^{18}$ H+/cm$^2$ data would exhibit the lowest magnetization. [6-9] Furthermore, the observed trend persists through the ferromagnetic to paramagnetic transition (Figure 4b).

We will now focus on the uniaxial anisotropy as a function of proton fluence. It is unlikely that damage to the CST crystals from increased proton irradiation is responsible for the observed reduction in magnetization because the trend between proton fluence and reduced magnetization is not proportional. Therefore, we examined the interatomic interaction of CST by investigating the changes in the magnetocrystalline anisotropy as a function of temperature and proton irradiation fluence. Here, we extracted the effective anisotropy constant, K$_u$, using the Stoner-Wohlfarth model shown in equation 1: [12,16-20]

$$\frac{2K_u}{M_{sat}} = \mu_0 H_{sat} \qquad (1)$$

Here, M$_{sat}$ is the saturation magnetization, H$_{sat}$ is the anisotropy field (i.e., the saturation field along the hard axis), and $\mu_0$ is the vacuum permeability. For clarity, the anisotropy field and the saturation magnetization are extracted from magnetic field vs. magnetization measurements performed in the H//ab direction. This allowed us to capture the magnitude of the magnetic field necessary to completely align the moments along an unpreferable direction. To estimate the onset of saturation along this direction, the second derivative of the first quadrant magnetization was taken with respect to the magnetic field and the field magnitude at the local minimum was taken as the saturation/anisotropy field. The multiplicative factor $\mu_0$ is simply unity in cgs units. The values of $H_{sat}$ (Figure 5a) and $K_u$ (Figure 5b) for pristine CST acquired from this estimation agree well with literature values. [16] In Figure 5b, $K_u$ is shown for all fluences as a function of temperature below $T_C$. Only temperatures below $T_C$ are selected as saturation is not achieved in the paramagnetic region. As expected, the overall values decrease with increasing temperature, but the fluence dependent behavior at each temperature closely mirrors that of the saturation magnetization with no simple linear relation between the two.

The anisotropy constant for the sample irradiated with 5 x $10^{16}$ $H^+/cm^2$ is consistently lower than that of the other proton fluences across all temperatures. Initially, it appears that this could be due to both a reduced $H_{sat}$ and $M_{sat}$, but a closer inspection of the former reveals a deeper phenomenon related to the temperature dependence of the anisotropy. In fact, a temperature-dependent $H_{sat}$ is direct evidence of a temperature-dependent anisotropic behavior in CST. [19,20] Moreover, the anisotropy field shows a slight proportionality with the proton fluence. As shown in Figure 5b, we observe that samples irradiated with more protons per unit area generally require a higher H//ab field to achieve saturation. This trend can be more observed more clearly at 20 K (Figure 5a, blue circles), with deviations at other temperatures likely resulting from measurement error.

We observe a decrease in $K_u$ for all fluences as the temperature is increased. This phenomenon has been documented in other vdW compounds such as $CrI_3$ and $Fe_3GeTe_2$ and is attributed to fluctuating local spin clusters around a macroscopic magnetization vector. [12,19,20] Hence, a lower value for $K_u$ in the irradiated samples might indicate an effect on the short-range interactions in this material. The CST anisotropy is stabilized by two factors, with the first being a slight distortion of the Te octahedra along the c-axis. The second factor is the Te p spin-orbit coupling through the dominant superexchange pathway in CST. [21-24] This Cr-Te-Cr superexchange is ferromagnetic, in agreement with the Goodenough-Kanamori rules, and dominates over the weak antiferromagnetic $Cr^{3+}$ direct exchange, the non-negligible interlayer interactions, and the 2nd and 3rd nearest neighbor intralayer interactions. [25] Therefore, changes in the anisotropy constant could be indirect evidence of modifications to the superexchange interactions present in proton irradiated CST, since it has been reported that the anisotropy of CST is sensitive to changes in both the superexchange interaction and electron-electron repulsions. [23]

In addition, though the $2^{nd}$ and $3^{rd}$ nearest neighbor interactions represent a superexchange pathway across two Te ligand atoms (Cr-Te-Te-Cr), they present a suitable environment for antiferromagnetic exchange couplings to exist in CST. [22, 23] If this is enhanced through proton irradiation, the antiferromagnetic exchange could account for the reduction in magnetization despite an unchanged $T_C$. The dependence of the magnetic properties on proton fluence remains unclear, however, and additional investigations are required to fully understand the nature of the exchange couplings present in CST.

We will now discuss the results obtained from EPR spectroscopy on CST as a function of proton fluence and temperature. The EPR spectra for all samples were collected at temperatures ranging from 13 K to 77 K in order to capture the ferromagnetic to paramagnetic phase transition in CST. EPR spectra as a function of proton fluence are shown at 13 K (Figure 6a), 35 K (Figure 6b), 50 K (Figure 6c), and 77 K (Figure 6d). These temperatures highlight the behavior of CST well into the ferromagnetic phase, then above $T_C$, and finally in the paramagnetic phase. We observe the expected signal from the $Cr^{3+}$ ion at all temperatures and all fluences with a g-value that ranges from 1.96109 - 1.9950 at 13 K depending on the fluence. [27-31] To extract the linewidth and (resonance) center field values, the data were fitted using a Lorentzian line shape function (dotted black line). In the spectrum of the pristine sample at 13 K, the $Cr^{3+}$ signal dominates, but further inspection shows yet another resonance that contributes to the spectrum, which is more readily observed in the spectra of the irradiated samples. This broad signal is centered at lower fields, and could not be completely captured by the available magnetic field range of our EPR spectrometer. Hence, further experiments at the high frequency EPR (>9 GHz) are required to pinpoint its origin.

This secondary signal is featured prominently even at 50 K, which is above the $T_C$. It has been previously reported that two-dimensional ferromagnetic correlations can persist in CST even up to room temperature. [15, 22, 32] Thus, it is possible that the existence of the secondary signal could be related to these correlations. However, at 77 K this secondary signal is still present but is greatly suppressed, with the spectra across all fluences being dominated by the $Cr^{3+}$ signal.

The temperature dependence of the g-value and peak-to-peak linewidth $H_{pp}$ as a function of fluence was also explored (Figure 7) for the $Cr^{3+}$ signal. The temperatures selected correspond to the temperature points from the magnetization vs. magnetic field isotherms. In general, the irradiated compounds show a narrowing of the $Cr^{3+}$ signal below $T_C$, but in the paramagnetic phase, there is a broadening effect. The g-values for all fluences except 5 x $10^{15}$ H$^+$/cm$^2$ behave similarly in the selected temperature range. Firstly, deviations in g-value from the expected Cr$^{3+}$ value are minimal as the temperature is raised from 13 K. However, at 40 K, there is a significant increase in the g-value for all fluences, with the g-value for some fluences even exceeding 2. This effect is also present in the 5 x $10^{15}$ H$^+$/cm$^2$ fluence, which generally displayed lower g-values than all other fluences below 40 K. It is likely that the fluence and temperature dependent behavior of the EPR spectral parameters is closely related to the behavior of the secondary signal. Moreover, due to its dominance of the spectra, it could affect the accuracy of the Lorentzian fits used to extract the parameters. Therefore, without further EPR experiments at high frequencies, it is difficult to discern the underlying cause behind the observed changes. Regardless, the fluctuations in H$_{pp}$ and g-values and the dominance of the secondary signal in the EPR spectra of the irradiated samples at temperatures below T$_C$ could be an indication of changes to the paramagnetic centers of CST, the local magnetic environments.

**Conclusions and outlook**

Exfoliable CST crystals show the versatility of the magnetic properties of vdW compounds upon proton irradiation as a function of proton irradiation. In this study, CST crystals irradiated with up to 1 x $10^{18}$ H$^+$/cm$^2$ were studied in a variety of temperature and magnetic field ranges. Through temperature and magnetic field-dependent magnetization, we observe a ferromagnetic phase below T$_C$ ~ 36 K for all fluences. Though the T$_C$ remains unchanged after proton irradiation, the saturation magnetization is decreased. Further analysis on the magnetic anisotropy of CST shows a temperature and fluence dependent effect, which could indicate variations in the exchange mechanisms of CST. While further investigations are required, we speculate that the reduced magnetization and uniaxial anisotropy could be a result of reformed short-range intralayer interactions involving the 1$^{st}$, 2$^{nd}$, and 3$^{rd}$ nearest neighbor interactions and the interlayer exchange.

Further evidence of this could be taken from the EPR spectra of the irradiated CST, which show a superposition of two signals. While the expected Cr$^{3+}$ signal is present for all samples, a secondary signal dominates the entire spectra of the samples at varying degrees depending on the temperature and fluence. Although the secondary signal could not be accurately captured by our current experiments, its presence, along with the changes in g-value and linewidth of the Cr$^{3+}$ signal, point to modifications in the magnetic correlations of proton irradiated CST. Overall, the case of proton irradiated CST presents an exciting environment to not only probe the origin of ferromagnetism in CST and other related compounds, but also to explore with methods of tuning their magnetic properties for nanoscale magnetic devices and spintronic applications. In effect, proton irradiation could be used in tuning the magnetic properties and magnetic interactions of exfoliable vdW materials that may have tremendous implications for information technology and space exploration.

## Acknowledgements

This material is based upon work supported by the National Science Foundation Graduate Research Fellowship Program under Grant No. 184874.1 Any opinions, findings, and conclusions or recommendations expressed in this material are those of the author(s) and do not necessarily reflect the views of the National Science Foundation. S.R.S. and H.I. acknowledge support from the NSF-DMR (Award No. 2105109). SRS acknowledges support from NSF-MRI (Award No. 2018067). M.L.K. acknowledge financial support from the National Science Foundation (MRI Grant CHE-1828744 and Grant CHE-1900237). L.M.M acknowledges the useful discussions with H. S. Nair. L.M.M and S.R.S acknowledge support from a UTEP start-up grant. L.M.M acknowledges the Wiemer Family for awarding Student Endowment for Excellence, and NSF-LSAMP Ph.D. Fellowship. This publication was prepared by S. R. Singamaneni and co-authors under the award number 31310018M0019 from The University of Texas at El Paso (UTEP), Nuclear Regulatory Commission. The Statements, findings, conclusions, and recommendations are those of the author(s) and do not necessarily reflect the view of the (UTEP) or The US Nuclear Regulatory Commission. Work at Brookhaven National Laboratory is supported by the U.S. DOE under Contract No. DESC0012704 (materials synthesis).

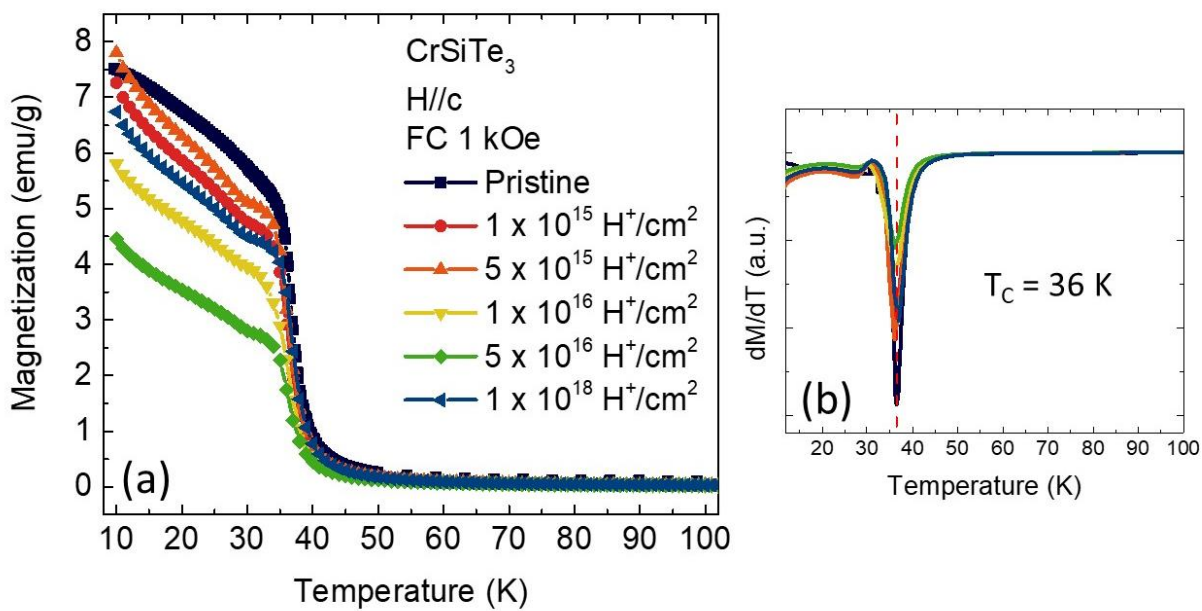

**Figure 1:** Magnetization vs. temperature (a) and dM/dT vs. temperature (b) shown as a function of fluence showing the ferromagnetic to paramagnetic transition at $T_c$ = 36 K for all samples

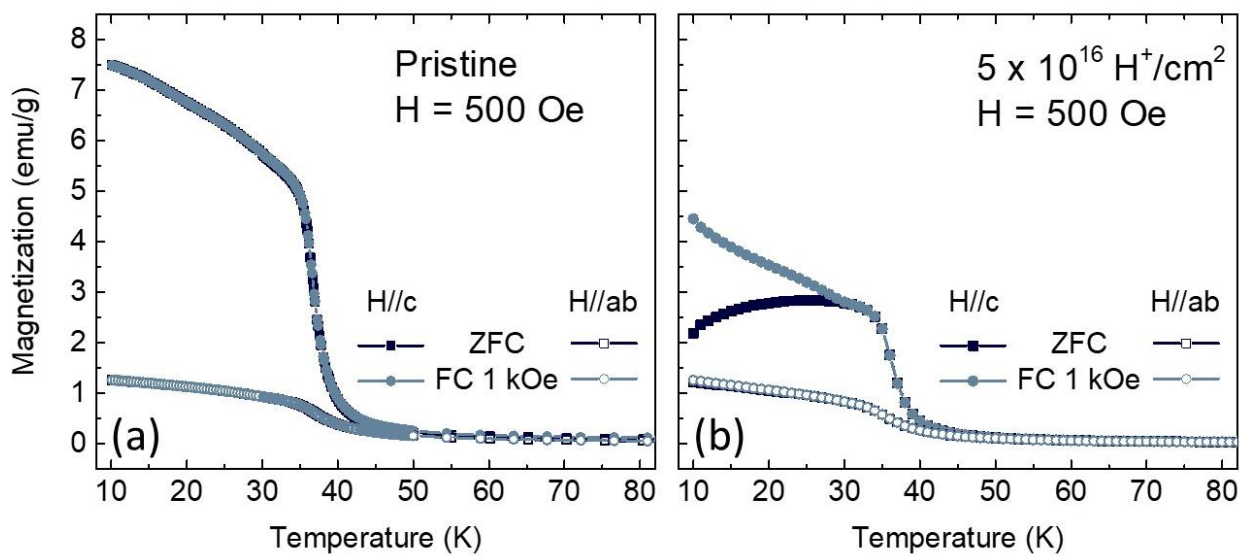

**Figure 2:** Anisotropic magnetization vs. temperature for both pristine (a) and irradiated (b) CST as a function of cooling field

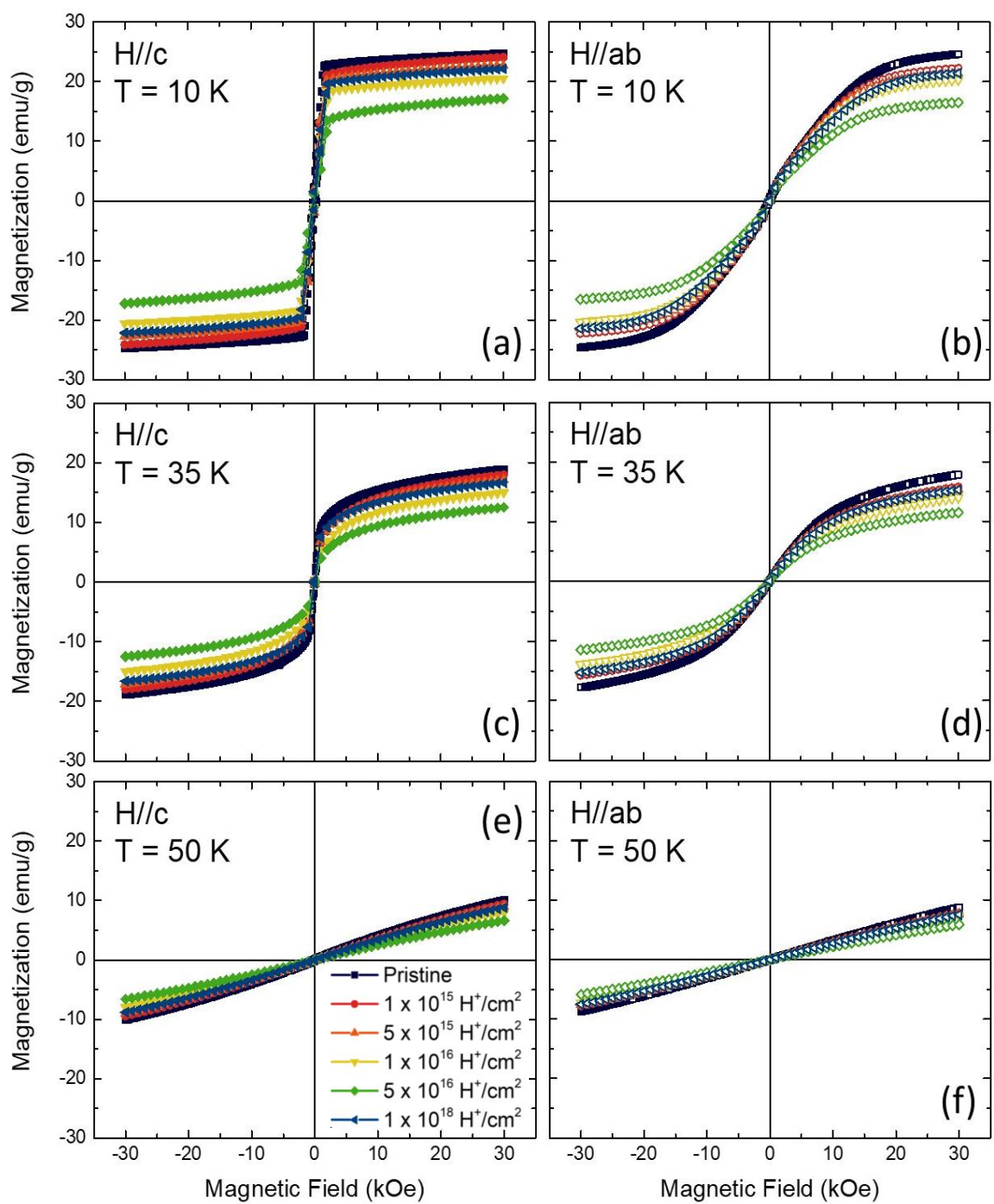

**Figure 3:** Magnetization vs. magnetic field as a function of fluence for H//c (a,c,e) and H//ab (b,d,f) at T = 10 K (a,b), 35 K (c,d), and 50 K (e,f)

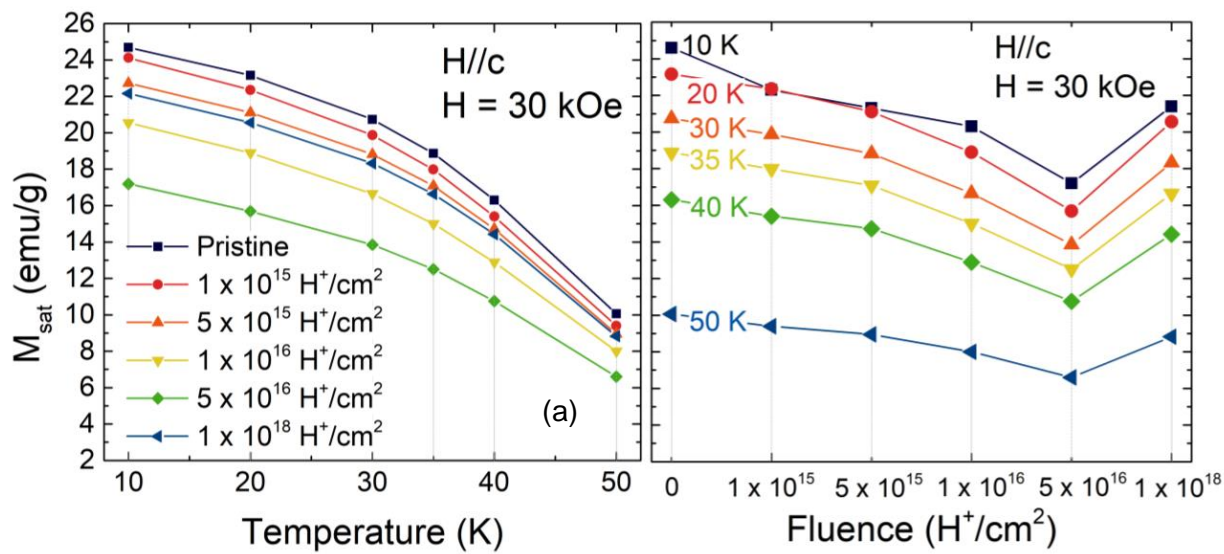

**Figure 4:** Saturation magnetization $M_{sat}$ vs. temperature as a function of fluence (a) and $M_{sat}$ vs. fluence at each temperature (b) showing non-monotonic relation between magnetization and proton irradiation

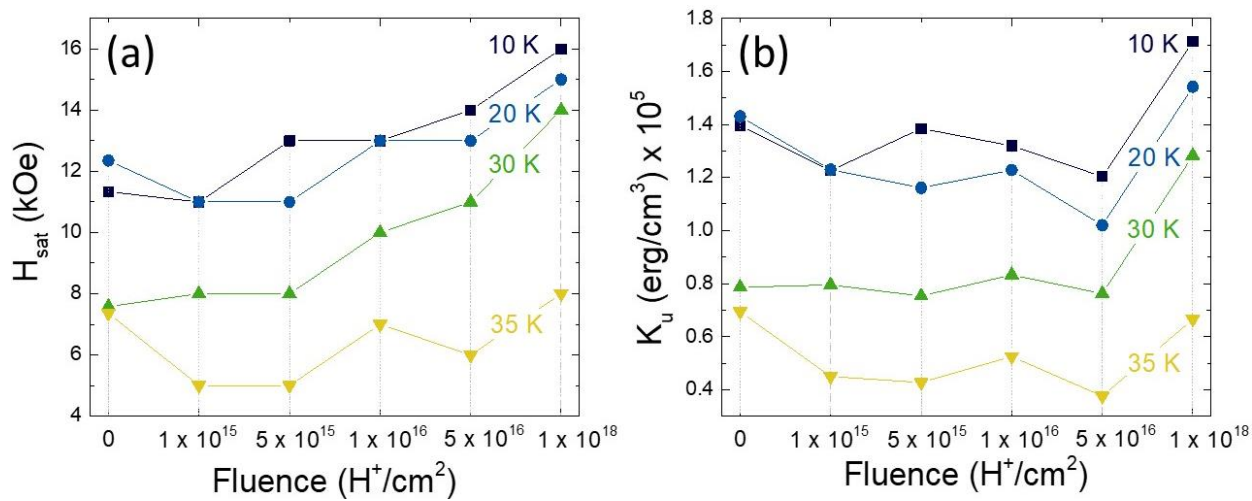

**Figure 5:** Anisotropy field $H_{sat}$ (a) and uniaxial anisotropy constant $K_u$ (b) vs. fluence as a function of temperature

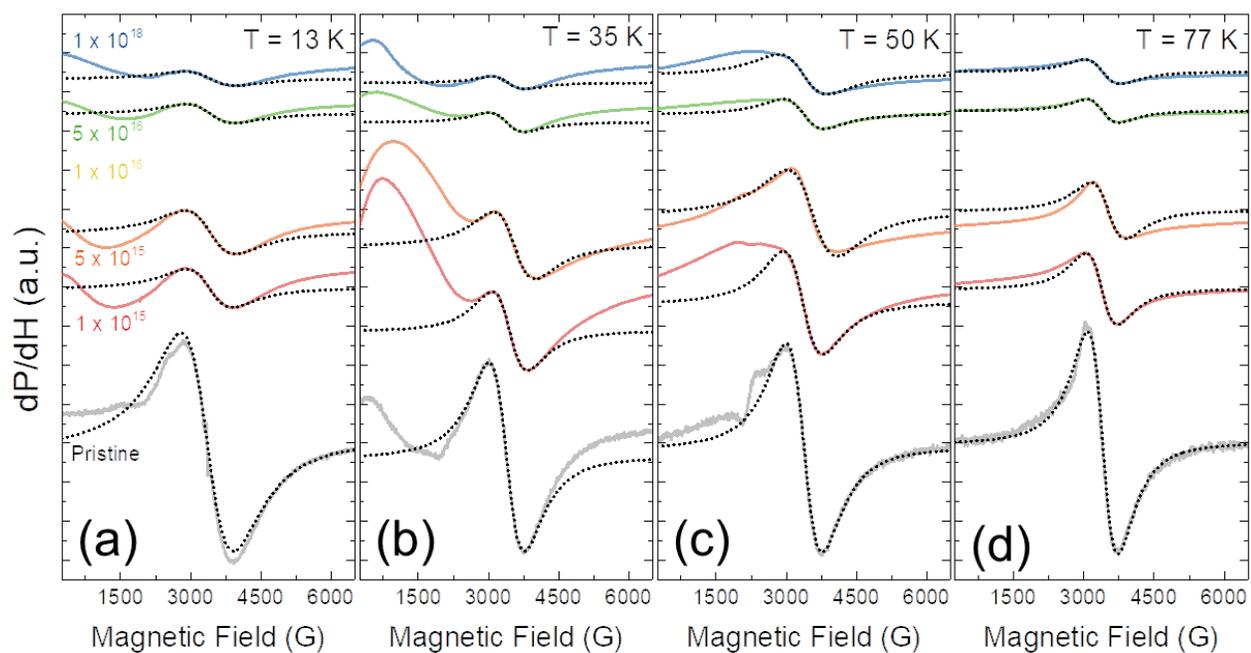

**Figure 6:** EPR spectra as a function of fluence at T = 13 K (a), 35 K (b), 50 K (c) and 77 K (d)

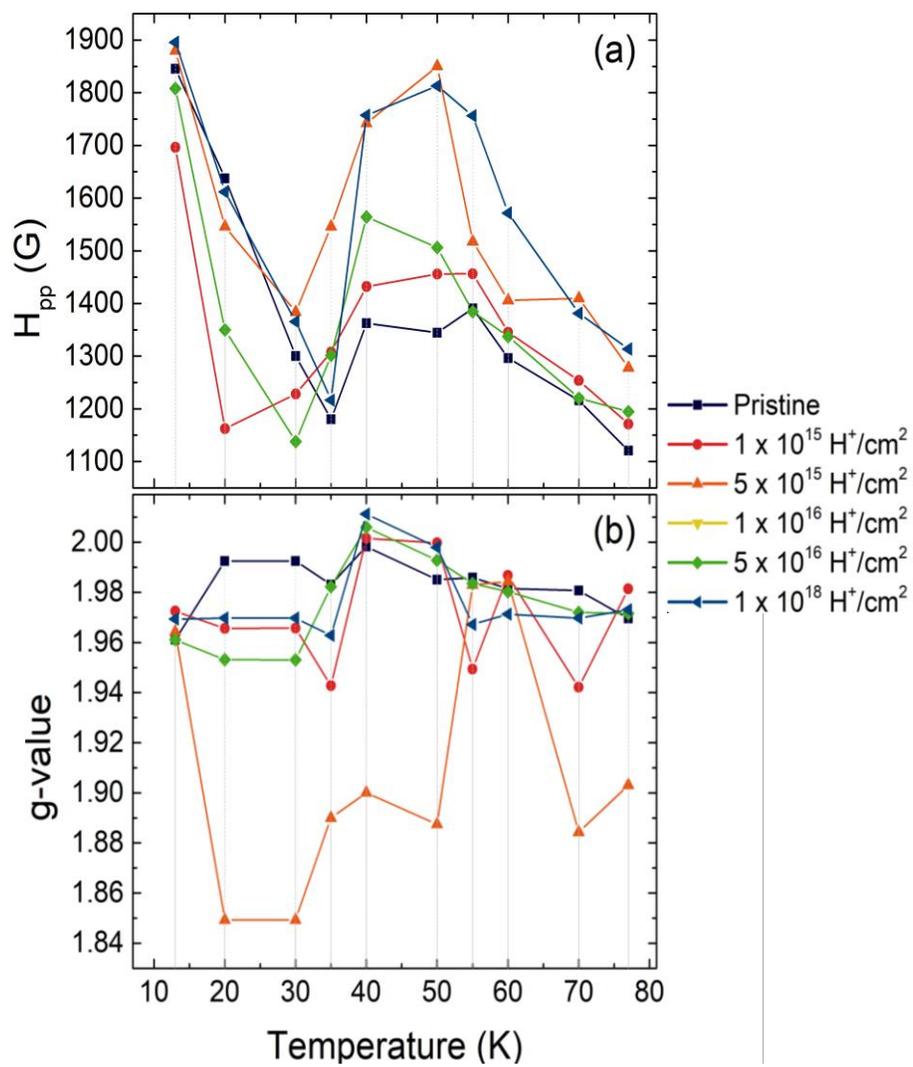

**Figure 7:** Peak-to-peak linewidth $H_{pp}$ (a) and g-value (b) vs. temperature as a function of proton fluence

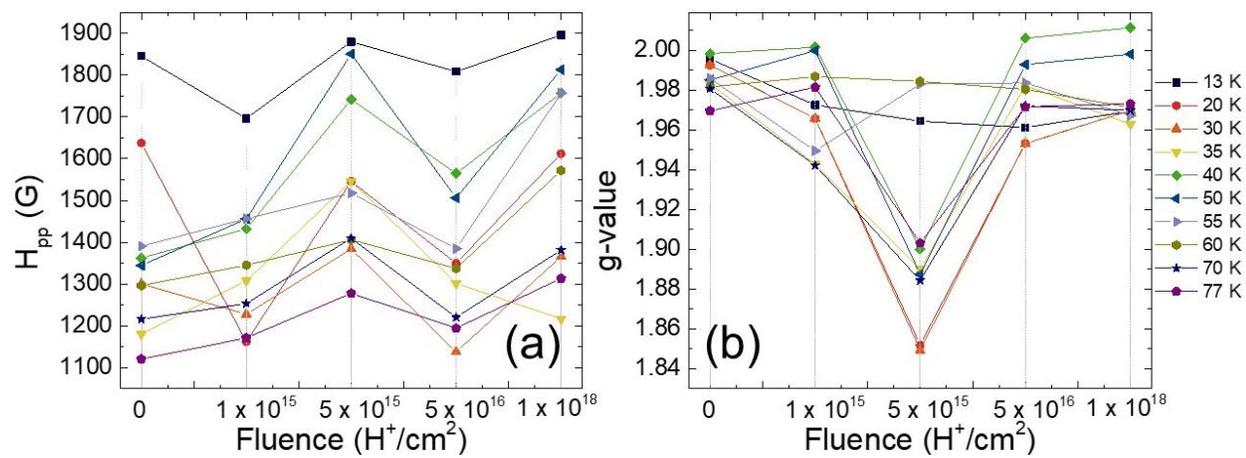

**Figure 8:** H$_{pp}$ (a) and g-value (b) vs. fluence as a function of temperature